\documentstyle[aps,twocolumn]{revtex}

\begin{document}
\draft

\title{St\"uckelberg Field Shiftting Quantization of Free-Particle on D-Dimensional Sphere}
\author{C.Neves(2,3) and C.Wotzasek(1)}
\address{(1)Instituto de F\'\i sica\\
Universidade Federal do Rio de Janeiro\\
21945-970, Rio de Janeiro, Brazil\\
(2)Centro Universitario de Barra Mansa\\
Barra Mansa, Rio de Janeiro, Brazil\\
(3)Departamento de F\'\i sica, ICE, Universidade Federal de Juiz de Fora,  
\\36036-330, Juiz de Fora, MG, Brasil,\\}
\date{\today}
\maketitle
\begin{abstract}
\noindent  In this paper we quantize the free-particle on a D-dimensional sphere in an unambiguous way by converting the second-class
constraint using St\"uckelberg field shiftting formalism. Further, we argument that this formalism is equivalent to the BFFT constraint
conversion method and show that the energy spectrum is identical to the pure
Laplace-Beltrami operator without additional terms arising from the curvature of the sphere. We work out the gauge symmetry generators with
results consistent with those obtained through the nonlinear implementation of the gauge symmetry.
\end{abstract}
\pacs{64.60.Cn,61.41.+e,64.75.+g}
\bigskip

\newpage

\section{Introduction}

The canonical quantization of the free-particle moving in a curved space is a fundamental theoretical problem that has
been investigated intensively over last decades in different settings\cite{DW,EG,OS,FH} but remains a controversial problem in the
literature. The relevance of this problem for the quantization on curvilinear surfaces
is well appreciated and its quantization has been studied both in the path integral and in the canonical approach. The quantum picture
however remains troubled by operator ordering ambiguities\cite{DW} and the results following
different approaches are not in complete agreement\cite{AB}. While most investigations have been done towards understanding the quantum nature directly from the
second-class formulation, a possible loophole to avoid problems would be the reformulation of the model as a gauge theory.

The proposal of this paper is the construction of a gauge invariant reformulation for the free point particle on the spherical surface
through the St\"uckelberg field shiftting formalism\cite{Stuckelberg}. This is possible after the nonlinear implementation of the
St\"uckelberg symmetry through the elimination of the Lagrange multiplier sector of the invariant theory.

The treatment of nonlinear systems as gauge theories was originally proposed by Kovner and Rosenstein \cite{KR}, using an analogy with QED
to disclose a symmetry hidden in the nonlinear sigma model (NLSM). An invariant version of this model was proposed by us\cite{NW} to explain
the results of \cite{KR} using the iterative constraint conversion approach\cite{IJMP}. Recently, different authors\cite{BGB,BN,BW} proposed distinct first-class versions for the spherical model using the BFFT formalism\cite{BFFT}.Due to the possibility of dealing with the nonlinear
constraint of the massive Yang-Mills theory through the constraint conversion technique, this problem has experienced a revival \cite{BBN,AW,KHP}. These works discuss the energy spectrum of the
collective mode of the theory and put under suspicion the result proposed by Adkins, Nappi and Witten (ANW) in Ref.\cite{ANW}.

It is worth mentioning that since the seminal work of Skyrme\cite{S} incorporating baryons in the NLSM low-energy description of the strong
interactions, the investigation of nonlinear theories has attracted much attention. The NLSM is a very useful model present in all
areas of physics. In condensed matter, for instance, it is used to describe systems ranging from anti-ferromagnetic spin-chains to certain
materials exhibiting fractional quantum Hall effect\cite{manu}. In lower dimensional physics, where it possess exact solution\cite{ABZ}, it
has became an important theoretical laboratory mainly due to its similarity to 4D nonabelian gauge theories with which it shares many
features such as renormalizability, asymptotic freedom, dynamical mass generation, confinement and topological excitations. It has also
been used in the theoretical investigation of the phenomenon of fractional spin and statistics in (2+1)D\cite{FW} and nonabelian
bosonization in (1+1)D\cite{EW}.

In the study of the static properties of nucleons, done by Adkins et al\cite{ANW}, a collective semi-classical expansion is performed by the
usual decomposition of the SU(2) matrix into the nonlinear sigma model action as,

\begin{equation}
U(r,t)=A(t)U(r)A(t)^{-1}
\end{equation}
where the matrix A(t) as $A(t)=a^o+ia.\tau$ satisfies the spherical constraint,

\begin{equation}
\label{00}
\phi_1 = a_ia_i - 1 = 0,\hspace{1cm} \mbox{with} \;\; i=0,1,2,3.
\end{equation}
The theory gets reduced to a nonlinear quantum mechanical model whose dynamics is governed by a Lagrangian dependent on
$a_i(t)$ and $\dot a_i(t)$ playing the roles of the particle's coordinate and velocity respectively. Similarly, the study of the fractional
spin and statistics in the context of the O(3) NLSM is reduced to that of the quantum rotor through the semi-classical separation of the
collective mode, reducing the problem to that of quantizing the spherical top. Recall that the spherical rotor\cite{FH,AB} is the paradigm of
the second-class constrained system with field dependent Dirac brackets\cite{PD}. Therefore the ambiguities resulting from the quantization
of this model affect the results above mentioned. This leads to the necessity of new studies that may eventually shed
some light over these questions.
\smallskip

\section{The spherical gauge model}

To  evidentiate the problem that affects the quantization process of nonlinear model, let us begin quantizing the system with the
Dirac's method for second-class constraints. A free point particle with unitary mass moving on a flat D+1-dimensional Euclidean space is
restricted to the D-spherical surface by the spherical constraint in configuration space

\begin{equation}
\label{05}
q_i q_i - R^2 = 0. 
\end{equation}
R represents the radius of the sphere and $q_i(t)$, i=1,2,...,D, are the particle's coordinates. The point
particle has its dynamic governed by the Lagrangian 

\begin{equation}
\label{06}
L = \frac 12 \dot q^2 + \lambda(q_i q_i - R^2 ).
\end{equation}
The corresponding Hamiltonian is,

\begin{equation}
\label{07}
H = \frac 12 p_i^2 - \lambda(q_i q_i - R^2 ).
\end{equation}
The constraint analysis reveals the presence of four second-class constraints,

\begin{eqnarray}
\label{A20}
\Omega_1 &=& \pi_\lambda\nonumber\\
\Omega_2 &=& q^2 - c\nonumber\\
\Omega_3 &=& p_i q_i\nonumber\\
\Omega_4 &=& p_i p_i - 2\lambda q_i q_i .
\end{eqnarray}

The geometrical meaning of $\Omega_2$ and $\Omega_3$ is transparent. The constraint $\Omega_2=0$ restrains the particle to move on the
D-sphere surface, while $\Omega_3$ means that the particle momentum remains tangent to nonlinear surface, without radial component during
the motion. The remaining constraints, $\Omega_1$ and $\Omega_4$, have no geometrical meaning and dynamical importance to the theory since
they are artifacts of constructing the Hamiltonian formalism from the Lagrangian using the Legendre transformation. It occurs because the
Lagrange multiplier $\lambda$, which enforces the nonlinear constraint in the Lagrangian formalism, is assumed to be an independent
dynamical variable. In this way, the Hamiltonian formalism yields these extra constraints to suppress the dynamics of $\lambda$ and
$p_\lambda$.

From the last condition of (\ref{A20}) the Lagrange multiplier could be computed explicitly as\cite{BGB},

\begin{equation}
\label{10}
\lambda = \frac 12 \frac{p^2}{q^2}.
\end{equation}
The particle's dynamic passes to be described by $H = \frac 12 p^2$ and two constraints
$(\Omega_2,\Omega_3)$. The symplectic structure on the physical phase space determined by these constraints is induced by
the Dirac brackets,

\begin{eqnarray}
\label{20}
\lbrace q_i, q_j\rbrace^* &=& \lbrace p_i, p_j\rbrace^* = 0,\nonumber\\
\lbrace q_i, p_j\rbrace^* &=& M_{ij},\nonumber\\
\lbrace p_i, p_j\rbrace^* &=& H_{ij}.
\end{eqnarray}
where 

\begin{eqnarray}
\label{matrix_M}
M_{ij}&=&\delta_{ij} - \frac {q_iq_j}{q^2},\nonumber\\
H_{ij}&=&\frac {(q_jp_i - q_ip_j)}{q^2}.
\end{eqnarray}

It may be stressed that the same results may be obtained from the simplified Lagrangean formulation with the proviso that the equation of motion of the eliminated variable must be mantained as a subsidiary condition to impart consistency in the canonical analysis. 
Next we present a proposal to express it as a gauge theory using the St\"uckelberg field shiftting formalism.

There are in the literature alternative methods to implement this proposal.  We quote the  BFFT\cite{BFFT} and
iterative\cite{NW,IJMP} methods that have attracted much attention in the literature. The BFFT conversion technique uses as many auxiliary variables
as the number of
second-class constraints\cite{BFFT}.  As mentioned, the analysis can be considerably simplified by eliminated the
multiplier sector of the phase space using the noninvariant character of the constraints\cite{davis}. The question
that seems to be of importance is related to the elimination of this sector before or after the constraint conversion.  The induced gauge
symmetry over the spherical model becomes realized nonlinearly or linearly respectively, leading to distinct consequences.  In the former
case, worked out in \cite{BGB}, the multiplier is eliminated before the BFFT procedure, based on its second-class character.  On the other
hand, without elimination of the multiplier sector, the gauge symmetry is linearly implemented by the Stuckelberg procedure.
Although we advocate the later procedure mostly because of its effectiveness and simplicity, we will show next that they indeed lead to
(canonically) equivalent results.
 
Let us consider the construction of the Wess-Zumino terms through St\"uckelberg mechanism,
\begin{eqnarray}
\label{A30}
L_{WZ}(\theta , \lambda) &=& L(\lambda - \frac 12\dot\theta) - L(\lambda)\nonumber\\
&=& \theta q_k \dot q_k .
\end{eqnarray}
An equivalent procedure using the
iterative conversion of the nonlinear constraints was given in \cite{NW}, whose gauge invariant Lagrangian was found to be

\begin{equation}
\label{50}
L = \frac 12 {\dot q_i}^2 + \theta(q.\dot q) + \lambda(q^2 - R^2),
\end{equation}
where $\theta$ is the WZ variable. The corresponding Hamiltonian, obtained reducing the Lagrangian (\ref{50}) to
first-order, is

\begin{equation}
\label{60}
H = \frac 12 p^2 + \frac 12 \theta^2 q^2 - \theta(q.p) - \lambda(q^2 - R^2).
\end{equation}
This theory has two chains of constraints whose primary members are,

\begin{eqnarray}
\label{70}
\phi_1=\pi_\lambda &\approx& 0,\nonumber\\
\psi_1=\pi_\theta &\approx& 0.
\end{eqnarray}
Since these constraints must to satisfy some integrability condition, it is required the presence of secondary constraints that are, respectively,

\begin{eqnarray}
\label{80}
\phi_2=q^2 - R^2 &\approx& 0,\nonumber\\
\psi_2=q.p -\theta q^2 &\approx& 0,
\end{eqnarray}
and no more constraints are generated by following Dirac's algorithm.
Although a naive inspection shows the presence of second-class constraints, the computation of the Dirac matrix shows the presence of two zero-modes, indicating the existence of a two distinct set of constraints.  One with two first-class constraints ($\varphi^{(1)}_k$) and other with two second-class constraints ($\varphi^{(2)}_k$), that are identified after a diagonalization of the Dirac matrix as,

\begin{eqnarray}
\label{RC1}
\varphi^{(1)}_1&=& \phi_1\nonumber\\
\varphi^{(1)}_2&=& \phi_2 - 2 \psi_1
\end{eqnarray}
and

\begin{eqnarray}
\label{RC2}
\varphi^{(2)}_1&=& \psi_1\nonumber\\
\varphi^{(2)}_2&=& \psi_2.
\end{eqnarray}

The elimination of the second-class sector is done via the Dirac bracket reduction, as usual.  It generates the following first-class Hamiltonian,

\begin{equation}
\label{RC3}
\bar{H}=\frac {1}{2} p_k \left( \delta_{km} - \frac{q_k q_m}{q^2}\right)p_m 
\end{equation}
where the reduced first-class constraints now read,

\begin{eqnarray}
\label{RC4}
\varphi^{(1)}_1\to \bar\varphi^{(1)}_1&=& \phi_1\nonumber\\
\varphi^{(1)}_2\to \bar\varphi^{(1)}_2&=& \phi_2.
\end{eqnarray}
Similarly, the original Poisson brackets are now mapped into Dirac brackets,

\begin{eqnarray}
\label{RC5}
\lbrace q_i,q_j\rbrace^* &=& 0\nonumber\\
\lbrace p_i,p_j\rbrace^* &=& 0\\
\lbrace q_i,p_j\rbrace^* &=& \delta_{ij}\nonumber.
\end{eqnarray}
Notice that the Dirac brackets for this partial reduction of constraints has a canonical structure.
This just reflects the result of the well known Maskawa-Nakashima theorem\cite{MN}.
This new Hamiltonian and sympletic structure define a pure first-class problem. By a simple inspection the correct equation of motion may be obtained from these objects. The symmetries transformations are generated by these constraints as $\delta_i {\cal O}= \varepsilon_i\lbrace {\cal O}, \bar\varphi^{(1)}_i\rbrace^* $ ($i=1,2$ and ${\cal O}=\lambda,q_k ,p_k$),

\begin{eqnarray}
\label{RC6}
\delta_1 \lambda &=& \varepsilon_1,\nonumber\\
\delta_1 q_i &=& \delta_1 p_i = 0,
\end{eqnarray}
and 

\begin{eqnarray}
\label{RC7}
\delta_2 \lambda &=& 0,\nonumber\\
\delta_2 q_i &=& 0,\\
\delta_2 p_i &=& - 2\varepsilon_2 q_i.\nonumber
\end{eqnarray}
Notice that the coordinates $q_i$ are null eigenvectors of the matrix $M_{ij}$, defined in (\ref{matrix_M}), acting as a phase space metric in the reduced Hamiltonian $\bar H$ in (\ref{RC3}).
To complete this discussion it is important to recall that the computation of the two set of constraints given in (\ref{RC1}) and (\ref{RC2}) was imperative to the development of this procedure. However, the splitting computation of the original set of constraints may be obscure. To avoid this problem a systematic alternative, based on the reduction of the set of constraints with the elimination of the superfluous constraints is next elaborate that will illuminate the full power of the St\"uckelberg formalism.

Let us recall that the theory in discussion (\ref{50}) is known to possess four constraints.  However, for systems with holonomous constraints imposed by Lagrange multipliers, some of these constraints only appear in the canonical process to eliminate the dynamics associated with the multiplier sector of variables.  It is usual practice to use an improved Hamiltonian obtained by eliminating the Lagrange multiplier sector {\it ab initio}.  This will keep only the meaningful geometrical constraints and simplify the analysis.
However recall that the equation of motion associated to the (eliminated) multiplier sector is mantained as a consistency condition to the canonical structure associated to the simplified Lagrangean formulation.
To eliminate the redundant constraints we proceed as follows.  The Euler-Lagrange equations for $\theta$ and $q_i$, are solved,

\begin{eqnarray}
\label{90}
q.\dot q &=& 0,\nonumber\\
{\ddot q}_i + \dot\theta q_i - 2\lambda q_i &=& 0,
\end{eqnarray}
respectively which determine the Lagrange multiplier as,
\begin{equation}
\label{100}
\lambda = - \frac {1}{2q^2}({\dot q}^2 - \dot\theta q^2).
\end{equation}
The arbitrariness present in the multiplier reflects the gauge freedom induced over the system.
Bringing this relation back into the WZ theory we find a new canonical structure given by the modified Hamiltonian,

\begin{equation}
\label{120}
{\tilde H} = \frac 1{2R^2} q^2p^2 -\frac {q^2}{R^2}(q.p)\theta + \frac 1{2R^2} \theta^2 (q^2)^2.
\end{equation}
and the first-class, strongly involutive constraint

\begin{equation}
\label{130}
\omega_1= q^2 - R^2 - 2\pi_\theta \approx 0,
\end{equation}
which has no time evolution since,

\begin{equation}
\label{135}
\dot\omega_1=\lbrace \omega_1, \tilde H\rbrace =0.
\end{equation}
On the other hand consistence with the first equation in (\ref{90}) requires $\pi_\theta$ to have no time evolution of its own,  
\begin{equation}
\label{137}
0=\dot\pi_\theta=\lbrace \pi_\theta , \tilde H\rbrace .
\end{equation}
This condition imposes a new constraint over the system as,

\begin{equation}
\label{138}
\omega_2=q.p - \theta q^2
\end{equation}
This canonical structure is similar to the one obtained in \cite{BGB} and identical to that given in \cite{NW2}, after a convenient interchange the WZ variables $(\theta \rightleftharpoons \pi_\theta)$ with a corresponding change of signs.

Note that after the elimination of the multiplier $\lambda$ some aspects of the model have changed. Here, $\pi_\theta$ is not a constraint but it is a canonical variable of the model that is absorbed by the nonlinear constraint, deforming the original spherical surface and destroying the constraint hierarchy given in (\ref{70}) and (\ref{80}). Consequently, after the exceeding constraints are eliminated, the remaining geometrical ones, and the gauge invariant model described by the Hamiltonian (\ref{120}) are equivalent to the original first-class system given in (\ref{60}). This reduced gauge invariant model has two first-class constraints $\omega_1$ and $\omega_2$,
that obey the strongly involutive algebra,

\begin{eqnarray}
\label{150}
\lbrace q^2 - R^2 - 2\pi_\theta,{\tilde H} \rbrace &=& 0,\nonumber\\
\lbrace q.p - \theta q^2, {\tilde H}\rbrace &=& 0,
\end{eqnarray}
which is in agreement with the issues of \cite{BGB} and \cite{NW2}. These first-class constraints generate the following infinitesimal gauge transformations on the canonical variables in the complete extended space: 

\begin{eqnarray}
\label{160}
\delta_1 q_i &=& \varepsilon_1\lbrace q_i,\omega_1\rbrace = 0,\nonumber\\
\delta_1 p_i &=& \varepsilon_1\lbrace p_i,\omega_1\rbrace = -2\varepsilon_1 q_i,\nonumber\\
\delta_1 \pi_\theta &=& \varepsilon_1\lbrace \pi_\theta,\omega_1\rbrace = 0,\nonumber\\
\delta_1 \theta &=& \varepsilon_1\lbrace \theta,\omega_1\rbrace = -2\varepsilon_1,
\end{eqnarray}
and
\begin{eqnarray}
\delta_2 q_i &=& \varepsilon_2\lbrace q_i,\omega_2\rbrace = \varepsilon_2 q_i,\nonumber\\
\delta_2 p_i &=& \varepsilon_2\lbrace p_i,\omega_2\rbrace = - \varepsilon_2 (p_i - 2\theta q_i),\nonumber\\
\delta_2 \pi_\theta &=& \varepsilon_2\lbrace \pi_\theta,\omega_2\rbrace = \varepsilon_2 q_i^2,\nonumber\\
\delta_2 \theta &=& \varepsilon_2\lbrace \theta,\omega_2\rbrace = 0,
\end{eqnarray}
that agrees with those obtained in \cite{KR,NW,BGB,NW2}.
The finite induced WZ gauge symmetries within the extended phase space are obtained from the gauge generating constraints by successive application on the
canonical and the extended phase space variables,
 
\begin{eqnarray}
\label{180}
q_i&\rightarrow& e^{\varepsilon_2}q_i,\nonumber\\
p_i&\rightarrow& e^{-\varepsilon_2}p_i + 2 q_i(\theta  e^{\frac{\varepsilon_2}{2}} - (\theta + \varepsilon_1) e^{-\frac{\varepsilon_2}{2}}),\nonumber\\
\theta&\rightarrow&\theta - 2\varepsilon_1,\nonumber\\
\pi&\rightarrow&\pi + (1 - e^{-\varepsilon_2})q^2.
\end{eqnarray}
Note that the group of transformations generated by the first-class constraints act nonlinearly on the extended phase space.

We stress that Kovner-Rosenstein's hidden symmetry is indeed an induced symmetry of the Wess-Zumino sector over the phase space of the theory.
This effect, as discussed above is clearly independent of the particular method of constraint conversion, being quite unique. Indeed the constraint $\omega_1$ in (\ref{130}) is immediately transformed into the KR constraint generator with a special value for $\pi_\theta$.  Interestingly, this also corresponds to a choice of initial condition in (\ref{137}). This is
revealed by gauge-fixing the WZ sector in such a way to recover the spherical constraint as the symmetry generator of the KR symmetry. Either way, this may be achieved by adding the gauge-fixing constraint,

\begin{equation}
\omega_3 = \pi_\theta
\end{equation}
to the set $\omega_1$ and $\omega_2$.  The $\Omega=\omega_1|_{\pi_\theta}$ constraint now plays the role of Gauss law generator for the KR symmetry in the
original phase space, under the Dirac bracket algebra generated by the second-class constraints $\omega_2$ and $\omega_3$.  This reduced algebra has the same canonical structure as in (\ref{RC5}) which is another illustration of the Maskawa-Nakashima theorem\cite{MN}.  The dynamics is controlled by the Hamiltonian (\ref{120}) which on the constraint shell  $\omega_i\approx 0$ reads,

\begin{equation}
H_{KR}=\frac {1}{2R^2} q^2 p_k \left( \delta_{km} - \frac{q_k q_m}{q^2}\right)p_m
\end{equation}
which is seen to be the one postulate in \cite{KR}.
This purely first-class Hamiltonian structure leads to the correct field equations under the induced Dirac bracket algebra.

To realize the quantization it is necessary to introduce a gauge fixing term in order to fix the first-class nature of the gauss law.
Choosing the  gauge condition as

\begin{equation}
\label{AA10}
\Psi = p_D = 0,
\end{equation}
which is the canonical momentum conjugate to the coordinate $q_D$ and removes the dynamic of this coordinate. The Poisson brackets between
the constraints $\Omega$ and $\Psi$ is

\begin{equation}
\label{AA20}
\lbrace\Omega,\Psi\rbrace = 2 q_D,
\end{equation}
and as $q_D\not=0$ on the spherical surface, they form a set of second-class constraints and the theory passes to have 2D remaining phase space variables. The Dirac brackets among the independent variables are

\begin{eqnarray}
\label{AA30}
\lbrace q_\alpha,q_\beta\rbrace^* &=& 0,\nonumber\\
\lbrace q_\alpha,p_\beta\rbrace^* &=& \delta_{\alpha\beta},\\
\lbrace p_\alpha,p_\beta\rbrace^* &=& 0,\nonumber
\end{eqnarray}
where $\alpha$ and $\beta$ represents the independent phase space variables. The noninvariant Hamiltonian in the reduced phase space is

\begin{equation}
\label{AA40}
H = \frac {1}{2R^2} p_\alpha g^{\alpha\beta}p_\beta,
\end{equation}
with the non-singular phase space metric,

\begin{equation}
\label{AA50}
g^{\alpha\beta} = \delta^{\alpha\beta} - \frac {q^\alpha q^\beta}{R^2}.
\end{equation}

This Hamiltonian formulation of the problem has also been found by Abdalla and Banerjee \cite{AB} following a purely
second-class approach to quantize the system.  In the sequel we follow Ref.\cite{AB} closely in order to find the
spectrum of the skyrmion.
Since this system is unconstrained the velocities obtained from the Hamilton's equation of motion for $q_\alpha$,

\begin{equation}
\label{AA55}
\dot q_\alpha = g^{\alpha\beta} p_\beta,
\end{equation}
can be obtained in an unambiguous form from the canonical momenta by inverting the equation above,

\begin{equation}
\label{AA56}
p_\beta= g_{\alpha\beta} \dot q_\alpha,
\end{equation}
where $g_{\alpha\beta}$ is the inverse of (\ref{AA50}), being given by

\begin{equation}
\label{AA57}
g_{\alpha\beta} = \delta_{\alpha\beta} + \frac {q_\alpha q_\beta}{R^2 - q^2}.
\end{equation}

In the quantization of nonlinear models the ordering of phase space fields cannot be neglected since the Dirac brackets are field dependent, as carried out in Ref.\cite{sugano}. Therefore, there arises an important question as how to one should settle the quantum Hamiltonian from its corresponding classical description. The answer resides on the preservation of the classical symmetries in the quantum scenario\cite{OS}. In this way the corresponding quantum Hamiltonian is uniquely determined. Based in the quantum process developed in Ref.\cite{OS} the quantization of the reduced nonlinear model (\ref{AA40}) is accomplished if the reduced Hamiltonian is replaced by the corresponding Laplace-Beltrami operator defined as

\begin{eqnarray}
\label{AA60}
\hat H &=& - \frac 12 g^{-1/2}\partial_\alpha g^{\alpha\beta}g^{1/2}\partial_\beta,\nonumber\\
&=& - \frac 12 (R^2 - q^2)^{-1/2}\partial_\alpha g^{\alpha\beta}(R^2 - q^2)^{1/2}\partial_\beta,
\end{eqnarray}
where $\partial_\alpha=\frac {\partial}{\partial q_\alpha}$ are the derivatives with respect to the D-dimensional curved space coordinates,
and $g$ is the determinant of the metric $g_{\alpha\beta}$ given by,

\begin{eqnarray}
\label{AA61}
det[g_{\alpha\beta}] &=& exp\; tr\; ln \left(\delta_{\alpha\beta} + \frac {a_\alpha a_\beta}{R^2 - q^2}\right),\nonumber\\
&=& exp \; tr \frac {q_\alpha q_\beta}{q^2} \ln\left(1 + \frac{q^2}{R^2 - q^2}\right),\nonumber\\
&=& \frac {R^2}{R^2 - q^2}.
\end{eqnarray} 
Due to this, the Hamiltonian operator (\ref{AA60}) is related with the angular momentum in the reduced space,

\begin{eqnarray}
\label{AA62}
L_{\alpha\beta} &=& q_\alpha p_\beta - q_\beta p_\alpha = -i \hbar (q_\alpha \partial_\beta - q_\beta \partial_\alpha),\\
L_{\alpha D} &=& q_\alpha p_D - q_D p_\alpha = -i \hbar q_D \partial_\beta = - i\hbar (R^2 - q^2)^{1/2}\partial_\beta,\nonumber
\end{eqnarray}
and therefore it is rewritten as,

\begin{equation}
\label{AA23}
\hat H = \sum_{\alpha\beta} \frac {L_{\alpha\beta}^2}{2R^2}.
\end{equation}
Thus, we find that the quantum Hamiltonian is the conventional Schr\"odinger operator without any extra curvature term. Consequently, the
energy spectrum reads,

\begin{equation}
\label{AA64}
E = \frac 1{2R^2}{l(l + D -1)},
\end{equation}
in agreement with the result obtained by others authors\cite{AB,KS,PODOLSKY,LANDAU}. 

At this stage it is interesting to put our result in a more realistic framework that might shed some light over the question. To this end we
focus our discussions on the {\it Skyrme model}. There $D=3$ and consequently, the energy spectrum (\ref{AA64}) becomes,

\begin{equation}
\label{AA65}
E = \frac 1{2R^2}{l(l + 2)},
\end{equation}
that agrees with the result proposed by ANW\cite{ANW}. This completes our discussion.

\section{Conclusion}

In summary, the gauge symmetry of the nonlinear model is induced by phase space extension methods using the St\"uckelberg field
shiftting constraint conversion, displaying the equivalence with the constraint conversion methods. Afterward the energy spectrum was obtained without
additional constant term arising from the curvature of the D-sphere. Subsequently the {\it Skyrme model} was considered to study in this
scenario and the energy spectrum was also obtained without extra terms.

To finish this section it is important to exchange some views about the reduction process for the multiplier sector: if it is reduced before
or after the constraint conversion leads to distinct realizations of the WZ symmetry. We have verified that the procedure of reduction
commutes with the constraint conversion process, leading to results canonically equivalents.   This seems to be of importance for the analysis of nonabelian second-class systems as gauge theories where quadratic
constraints are intrinsically defined. Finally, it becomes clear that the question regarding the construction of the generators of the WZ
gauge symmetry cannot be tackled from this approach, in the sense that there is no plausible argument that favors any of the constraints as
the leader of the constraint chain. If this question becomes an important issue for the analysis of the problem at hand then the use of the
nonabelian BFFT method or the iterative constraint process seems unavoidable.\\

\noindent {\bf Acknowledgements}: This work is supported in part by CNPq, CAPES, FINEP and FUJB, Brazilian reasearch agencies.

\end{document}